\documentclass[twocolumn,epjc3]{svjour3}

\RequirePackage{graphicx}

\RequirePackage{amsmath,amsfonts,amssymb}
\RequirePackage[numbers,sort&compress]{natbib}
\RequirePackage[colorlinks,citecolor=blue,urlcolor=blue,linkcolor=blue]{hyperref}
\RequirePackage{dsfont}

\usepackage[utf8]{inputenc}

\journalname{Eur. Phys. J. C}

\begin{document}
\title{Quantum mechanical look at the radioactive-like decay of metastable dark energy}

\author{Marek Szyd{\l}owski\thanksref{oauj,csrc,e-ms}
\and
Aleksander Stachowski\thanksref{oauj,e-as}
\and
Krzysztof Urbanowski\thanksref{ifuz,e-ku}}

\thankstext{e-ms}{marek.szydlowski@uj.edu.pl}
\thankstext{e-as}{aleksander.stachowski@doctoral.uj.edu.pl}
\thankstext{e-ku}{K.Urbanowski@if.uz.zgora.pl}

\institute{Astronomical Observatory, Jagiellonian University, Orla 171, 30-244 Krakow, Poland \label{oauj}
\and
Mark Kac Complex Systems Research Centre, Jagiellonian University, {\L}ojasiewicza 11, 30-348 Krak{\'o}w, Poland \label{csrc}
\and
Institute of Physics, University of Zielona G{\'o}ra, Prof. Z. Szafrana 4a, 65-516 Zielona G{\'o}ra, Poland \label{ifuz}}

\date{Received: date / Accepted: date}

\maketitle

\begin{abstract}
We derive the Shafieloo, Hazra, Sahni and Starobinsky (SHSS) phenomenological formula for the radioactive-like decay of metastable dark energy directly from the quantum mechanics principles. For this aim we use the Fock-Krylov theory of quantum unstable states. We obtain deeper insight on the decay process as having three basic phases: the phase of radioactive decay, the next phase of damping oscillations, and finally the phase of power law decaying. We consider the cosmological model with matter and dark energy in the form of decaying metastable dark energy and study its dynamics in the framework of non-conservative cosmology with an interacting term determined by the running cosmological parameter. We study cosmological implications of metastable dark energy and estimate the characteristic time of ending of the radioactive-like decay epoch as 22296 of the present age of the Universe. We also confront the model with astronomical data which show that the model is in good agreement with the observations. Our general conclusion is that we are living in the epoch of the radioactive-like decay of metastable dark energy which is a relict of the quantum age of the Universe.
\end{abstract}

\section{Introduction}

We follow Krauss and Dent's paper and apply the Fock-Krylov theory of quantum unstable states to analyse a cosmological scenario with decaying dark energy \cite{Krauss:2007rx,fock,Fock:1978fqm,khalfin,Fonda:1978dk}. For this purpose we extend the Shafieloo, Hazra, Sahni and Starobinsky (SHSS) model of metastable dark energy with radioactive-like decay \cite{Shafieloo:2016bpk} and we give physical motivation arising directly from quantum mechanics for phenomenological formulas for SHSS model of the dark energy. We replace the radioactive, classical physics constant decay rate by the decay rate derived using the Fock-Krylov theory of quantum unstable states.

As a result we obtain a logistic-type radiative decay of dark energy, which is proceeded by the much slower decay process than the radioactive one known as the quantum Zeno effect. Within such an approach we find the energy of the system in the unstable state and the decay rate. The rigorous results show that these quantities both are time dependent. We find the exact analytical expression for them assuming that the density of energy distribution, $\omega (E)$, in the unstable state has the Breit-Wigner form. Using these results we also find late times asymptotic expression of these quantities. Then we assume that the dark energy density decays and that this is a quantum process. Starting from these assumptions we use the derived decay rate to analyze decay process of the dark energy density.

We study cosmological implications of a derived formula for decaying dark energy in the framework of the flat FRW cosmology. We find an extension of the standard cosmological model in the form of an interacting cosmology in which a conservation condition for the energy momentum tensor is not conserved due to an interaction between the dark energy and dark matter energy transfer. We show how the decay of the running lambda term can solve the cosmological constant problem and how it can modify the canonical scaling law of energy density for dark matter. We also test the model by astronomical observations. Our statistical analysis gives the best fit values of density parameters for each component of decaying vacuum of dark energy. Testing model with observational data we have found that dark energy can decay in three distinguished ways: exponentially, by damping oscillation and in power-law decay. We show that the main contribution to the decay of metastable vacuum is the dark energy decay of an exponential type and this type of decay dominates up to $22296\times T_0$, where $T_0$ is the present age of the Universe. Our calculations show that the exponential decay has only an intermediate character and will be replaced in the future evolution of the Universe by an oscillation decay and decay of $1/t^2$ type. From the estimation of model parameters we obtain that the decay half life should be much larger than the age of the Universe.

Today modern cosmology has a methodological status of some effective theory, which is described very well by current astronomical observations in terms of dark matter and dark energy. However, there are many open problems related to unknown nature of dark energy. The cosmological parameter is a good effective description of the accelerating phase of the current universe but we do not understand why the today value of this parameter is so small in comparison with its value in the early universe.

We look for an alternative cosmological model to supersede the $\Lambda$CDM model, the present standard cosmological model. Our main motivation is to find a solution of the cosmological constant problem. In this paper, the proposition of the solution of the problem of the cosmological constant parameter assumes that the vacuum energy is given by the fundamental theory \cite{Polchinski:2006gy}. We assume quantum mechanics as a fundamental theory, which determines cosmological parameters and explain how cosmological parameters change during the cosmic evolution. The discussion about the cosmological constant problem is included in papers \cite{Polchinski:2006gy,Weinberg:1988cp,Carroll:1991mt,Dolgov:1997za,Sahni:1999gb,Straumann:1999ia,Weinberg:2000yb,Carroll:2000fy,Rugh:2000ji,Padmanabhan:2002ji, Yokoyama:2003ii,Sarkar:2005bc,Copeland:2006wr,Szydlowski:2015bwa}.

Krauss and Dent \cite{Krauss:2007rx} analyzed properties of the false vacuum state form the point of view of the quantum theory of decay processes. They assumed that the decay process of metastable vacuum is a quantum decay process realized as the transition from the state corresponding to the metastable (false) vacuum state to the state corresponding to the lowest energy of the Universe (that is to the true vacuum state) and thus that this process can be described using standard quantum formalism usually used to describe decay of excited atomic levels or unstable particles. They used the Fock-Krylov theory of quantum unstable states \cite{fock,Fock:1978fqm,khalfin,Fonda:1978dk}. One of the famous results of this theory is the proof that quantum unstable systems cannot decay exponentially at very late times and that in such a late time regime any decay process must run slower than any exponentially decreasing function of time \cite{khalfin}. A model calculations show that survival probability exhibits inverse power law behavior at these times. Krauss and Dent \cite{Krauss:2007rx} analyzing a false vacuum decay pointed out that in an eternal inflation, many false vacuum regions can survive up to the times much later than times when the exponential decay law holds. They formulated the hypothesis that some false vacuum regions survive well up to the crossover time $T$ or later, where the crossover time, $T$, is the time when contributions of the exponential and late time non-exponential parts of the survival probability are of the same order. They gave a simple explanation of such an effect. It may occur even though regions of false vacua by assumption should decay exponentially, gravitational effects force space in a region that has not decayed yet to grow exponentially fast. Such a cosmological scenario may be realized if the lifetime of the metastable vacuum state or the dark energy density is much, much shorter than the age of the universe. It should be of order of times of the age of the inflationary stage of the Universe. 

The possibility that our Universe (or some regions in our Universe) were able to survive up to times longer that the crossover time $T$ should be considered seriously was concluded from Krauss and Dent's analysis \cite{Krauss:2007rx}. This is impossible within the standard approach of calculations of decay rate $\Gamma$ for decaying vacuum state \cite{Patrascioiu:1981vv,Coleman:1977py,Callan:1977pt,Coleman:1980aw,weinberg}. Calculations performed within this standard approach cannot lead to a correct description of the evolution of the Universe with false vacuum in all cases when the lifetime of the false vacuum state is such short that its survival probability exhibits an inverse power-law behavior at times comparable with the age of the Universe. This conclusion is valid not only when the dark energy density and its late time properties are related to the transition of the Universe from the false vacuum state to the true vacuum but also when the dark energy is formed by unstable "dark particles". In both cases the decay of the dark energy density is the quantum decay process and only the formalism based on the Fock-Krylov theory of quantum unstable states and used by Krauss and Dent \cite{Krauss:2007rx} is able to describe correctly such a situation. Note that Landim and Abdalla build a model of metastable dark energy, in which the observed vacuum energy is the value of the scalar potential at the false vacuum \cite{Landim:2016isc}.

Models with metastable dark energy have been recently discussed in the context of the explanation of the $H_0$ tension problem \cite{DiValentino:2017rcr}. Our model is a quantum generalization of Shafieloo et al.'s model\cite{Shafieloo:2016bpk} and contains a phase of radioactive-like decay valid in the context of solving this problem. Shafieloo et al. considered three different ways of dark energy decay. In our paper, we investigate the second way of the decay into dark matter. The models of the decay of the dark energy analyzed in \cite{Shafieloo:2016bpk} can be a useful tool for testing numerically decay processes discussed in \cite{Krauss:2007rx} and for analyzing properties of the decaying dark energy at times $t > T$. Namely, Shafieloo et al. \cite{Shafieloo:2016bpk} analyzed properties of the model of the time evolution of the dark energy. Their model assumes a ``radioactive decay'' scheme for decaying dark energy in which the present value of the dark energy density, $\rho_\text{DE}(t_{0})$, is related to its value at an earlier instant of time, $\rho_\text{DE}(t)$, by
 \begin{align}
 \rho_\text{DE}(t) &= \rho_\text{DE}(t_{0})\,\times\,\exp\,[- \Gamma\,(t-t_{0})]\equiv \rho_\text{DE}(t-t_{0}),\label{s-2-1}
 \end{align}
where the only free parameter is the decay rate $\Gamma$. Shafieloo et al. \cite{Shafieloo:2016bpk} derived this equation from the fundamental equation of the theory of radioactive decays
\begin{equation}
\dot{\rho}_\text{DE}(t) = -\,\Gamma\,\rho_\text{DE}(t), \label{s-2-2}
\end{equation}
(see eqs.~(2.1) and (2.2) in \cite{Shafieloo:2016bpk}). These equations are known from the Rutherford theory of decay of radioactive elements. Rutherford deriving these equations assumed that the number decaying radioactive elements at a given instant of time is proportional to a number of these elements at this moment of time \cite{Rutherford:1900a,Rutherford:1900b,Rutherford:1902a,Rutherford:1902b} as it is done in eq.~(\ref{s-2-2}). So the Rutherford's equations and thus also eqs.~(\ref{s-2-1})--(\ref{s-2-2}) are the classical physics equations.

In the context of equations (\ref{s-2-1})--(\ref{s-2-2}) one may ask what $\rho_\text{DE}(t)$ is built from that decays according to radioactive decay law? For physicists the only reasonable explanation for this problem is the assumption that $\rho_\text{DE}(t)$ describes the energy of an extremely huge number of particles occupying a volume $V_{0}$ at the initial instant of time $t_{0}$ and decaying at later times. Of course when such a particles can be considered as the classical particles then this process can be described using classical radioactive decay law. Unfortunately process of the creation of the Universe is not a classical physics process but it is a quantum process and particles or states of the system created during such a process exhibit quantum properties and are subject to the laws of quantum physics. The same concerns $\rho_\text{DE}(t)$ generated by quantum fluctuations or excitations of a quantum scalar field, which can be described as excited metastable states of this field and the process of their decay is the quantum process. Therefore as the quantum decay process it exhibits at late times completely different properties than the classical radioactive decay process what it was pointed out by Krauss and Dent. Simply, if $\rho_\text{DE}(t)$ is related to the extremely huge number of metastable states (excitations of the scalar field or its fluctuations) generated at $t_{0}$ in a volume $V_{0}$, it is very likely that many of them can be found undecayed at times longer than then the crossover time $T$. All this suggests that eqs.~(\ref{s-2-1}), (\ref{s-2-2}) may not be used when one wants to describe such a processes.

It seems that a reasonable way to make these equations suitable for description of quantum decay processes is to replace the quantity (a decay rate) $\Gamma$ appearing in eqs.~(\ref{s-2-1}) and (\ref{s-2-2}) by a corresponding decay rate derived using the quantum theory of unstable systems. The decay rate $\Gamma$ used in eqs.~(\ref{s-2-1}) and (\ref{s-2-2}) is constant in time but the decay rate derived within the quantum theory is constant to a very good approximation only at the so-called ``canonical decay regime'' of times $t$ (that is when the quantum decay law has the exponential form, i.e. when $t < T$) and at times $t$ much later than $T$ it tends to zero as $1/t$ when time $t$ tends to infinity (see, e.g, \cite{ku-pra}). This means that the decay process of a quantum unstable system is slower and slower for sufficiently late time which was also pointed out in \cite{Krauss:2007rx}.
This and other properties of the quantum decay process seem to be important when considering the cosmological inflationary as well late time (much later than the inflationary regime of times) processes including processes of transition of the dark energy density from its early time extremely large values to its present small value. Therefore we need quantities characterizing decay processes of quantum unstable systems.

The paper is organized as follows. Section 2 contains a brief introduction to the problems of unstable states and description of quantities characterizing such a states, which are used in next sections. In section 3 we analyze a possibility to describe metastable dark energy considering it as quantum unstable system. Section 4 contains a discussion of cosmological equations with decaying dark energy according to the quantum mechanical decay law and results of numerical calculations presented in graphical form. In section 5 we present statistical analysis. Section 6 contains conclusions.

\section{Preliminaries: Quantum unstable states}

 Properties of quantum unstable systems are characterized by their survival probability (decay law). The survival probability can be found knowing the initial unstable state $|\phi\rangle \in {\cal H}$, (${\cal H}$ is the Hilbert space of states of the considered system) of the quantum system, which was prepared at the initial instant $t_{0}$. The survival probability, ${\cal P}(t)$, of this state $|\phi\rangle$ decaying in vacuum equals
$
{\cal P}(t) = |A(t)|^{2},
$
where $A(t)$ is the probability amplitude of finding the system at the
time $t$ in the rest frame ${\cal O}_{0}$ in the initial unstable state $|\phi\rangle$,
$
A(t) = \langle \phi|\phi (t) \rangle .
$
and $|\phi (t)\rangle$ is the solution of the Schr\"{o}dinger equation
for the initial condition $|\phi (t_{0}) \rangle = |\phi\rangle$, which has the following form
 \begin{equation}
i \hbar \frac{\partial}{\partial t} |\phi (t) \rangle = \mathfrak{H} |\phi (t)\rangle. \label{Schrod}
\end{equation}
Here $|\phi \rangle, |\phi (t)\rangle \in {\cal H}$, and $\mathfrak{H}$ denotes the total self-adjoint Hamiltonian for the system considered. The spectrum of $\mathfrak{H}$ is assumed to be bounded from below: $E_{\text{min}} > - \infty$ is the lower bound of the spectrum $\sigma_{c}(\mathfrak{H}) = [E_{\text{min}}, +\infty) $ of $\mathfrak{H}$). Using the basis in ${\cal H}$ build from normalized eigenvectors $|E\rangle,\;\ E\in \sigma_{c}(\mathfrak{H})$ of $\mathfrak{H}$ and using the expansion of $|\phi\rangle$ in this basis one can express the amplitude $A(t)$ as the following Fourier integral
\begin{equation}
A(t) \equiv A(t - t_{0}) = \int_{E_{\text{min}}}^{\infty} \omega(E)\;
e^{\textstyle{-\,\frac{i}{\hbar}\,E\,(t - t_{0})}}\,d{E},
\label{a-spec}
\end{equation}
where $\omega(E) = \omega(E)^{\ast}$ and $\omega(E) > 0$ (see: \cite{fock,Fock:1978fqm,Fonda:1978dk}). Note that from the normalization condition ${\cal P}(0) \equiv |A(0)|^{2}=1$ it follows that there must be $ \int_{E_{\text{min}}}^{\infty} \omega(E)\; dE = 1$, which means that in the case of unstable states $\omega (E)$ is an absolutely integrable function. The consequence of this property is the conclusion following from the Riemann--Lebesgue lemma: There have to be $|A(t)| \to 0$ as $t \to \infty$. All these properties are the essence of the so-called Fock--Krylov theory of unstable states \cite{fock,Fock:1978fqm,Fonda:1978dk}. So within this approach the amplitude $A(t)$, and thus the decay law ${\cal P}(t)$ of the unstable state $|\phi\rangle$, are completely determined by the density of the energy distribution $\omega(E)$ for the system in this state \cite{fock,Fock:1978fqm} (see also \cite{khalfin,Fonda:1978dk,Kelkar:2010qn,muga,muga-1,calderon-2,Giraldi:2015a}. (This approach is also applicable in Quantum Field Theory models \cite{Giacosa:2011xa,goldberger}).

Note that in fact the amplitude $A(t)$ contains information about the decay law ${\cal P}(t)$ of the state $|\phi\rangle$, that is about the decay rate ${\it\Gamma}_{\phi}$ of this state, as well as the energy ${E}_{\phi}$ of the system in this state. This information can be extracted from $A(t)$. It can be done using the rigorous equation governing the time evolution in the subspace of unstable states, ${\cal H}_{\parallel} \ni |\phi\rangle_{\parallel} \equiv |\phi \rangle$. Such an equation follows from Schr\"{o}dinger equation (\ref{Schrod}) for the total state space ${\cal H}$.

Using Schr\"{o}dinger equation (\ref{Schrod}) one finds that within the problem considered
 \begin{equation}
i \hbar \frac{\partial}{\partial t}\langle\phi |\phi (t) \rangle = \langle \phi|\mathfrak{H} |\phi (t)\rangle. \label{h||1}
\end{equation}
From this relation one can conclude that the amplitude $A(t)$ satisfies the following equation
\begin{equation}
i \hbar \frac{\partial A(t)}{\partial t} = h(t)\,A(t), \label{h||2}
\end{equation}
where
\begin{equation}
h(t) = \frac{\langle \phi|\mathfrak{H} |\phi (t)\rangle}{A(t)} \equiv \frac{i\hbar}{A(t)}\,\frac{\partial A(t)}{\partial t} \label{h(t)}
\end{equation}
and $h(t)$ is the effective Hamiltonian governing the time evolution in the subspace of unstable states ${\cal H}_{\parallel}= \mathbb{P} {\cal H}$, where
$\mathbb{P} = |\phi\rangle \langle \phi|$ (see \cite{ku-pra} and also \cite{Urbanowski:2006mw,ku-2009} and references therein).
The subspace ${\cal H} \ominus {\cal H}_{\parallel} = {\cal H}_{\perp} \equiv \mathbb{Q} {\cal H}$ is the subspace of decay products. Here $\mathbb{Q} = \mathbb{I} - \mathbb{P}$. One meets the effective Hamiltonian $h(t)$ when one starts with the Schr\"{o}dinger equation for the total state space ${\cal H}$ and looks for the rigorous evolution equation for a distinguished subspace of states ${\cal H}_{||} \subset {\cal H}$ \cite{ku-pra,Giraldi:2015a}. In general $h(t)$ is a complex function of time and in the case of ${\cal H}_{\parallel}$ of dimension two or more the effective Hamiltonian governing the time evolution in such a subspace it is a non-hermitian matrix $H_{\parallel}$ or non-hermitian operator. There is
\begin{equation}
h(t) = E_{\phi}(t) - \frac{i}{2} {\it\Gamma}_{\phi}(t), \label{h-m+g}
\end{equation}
and
$
E_{\phi}(t) = \Re\,[h(t)],\; {\it\Gamma}_{\phi}(t) = -2\,\Im\,[h(t)],
$
are the instantaneous energy (mass) $E_{\phi}(t)$ and the instantaneous decay rate, ${\it\Gamma}_{\phi}(t)$ \cite{ku-pra,Urbanowski:2006mw,ku-2009}. (Here $\Re\,(z)$ and $\Im\,(z)$ denote the real and imaginary parts of $z$ respectively). The quantity ${\it\Gamma}_{\phi}(t) = -2\,\Im\,[h(t)]$ is interpreted as the decay rate because it satisfies the definition of the decay rate used in quantum theory: $\frac{{\it\Gamma}_{\phi}(t)}{\hbar} \stackrel{\rm def}{=} \,-\,\frac{1}{{\cal P}(t)}\,\frac{\partial {\cal P}(t)}{\partial t}$. Using (\ref{h(t)}) it is easy to check that
\begin{multline}
\frac{{\it\Gamma}_{\phi}(t)}{\hbar} \equiv
- \frac{1}{{\cal P}(t)} \frac{\partial {\cal P}(t)}{\partial t} =
- \frac{1}{|A(t)|^{2}}\;\frac{\partial |A(t)|^{2}}{\partial t} \\ 
\equiv - \frac{2}{\hbar}\,\Im\,[h(t)]. \label{G-equiv}
\end{multline}

The use of the effective Hamiltonian $h(t)$ leads to the following form of the solutions of eq.~(\ref{h||2})
\begin{equation}
A(t) = e^{\textstyle{-i\,\frac{t}{\hbar}\,\overline{h(t)}}} \equiv e^{\textstyle{-i\,\frac{t}{\hbar}\,(\overline{E_{\phi}(t)} - \frac{i}{2}\overline{{\it\Gamma}_{\phi}(t)})}}, \label{A-h}
\end{equation}
where $\overline{h(t)}$ is the average effective Hamiltonian $h(t)$ for the time interval $[0,t]$: $\overline{h(t)} \stackrel{\rm def}{=} \frac{1}{t}\,\int_{0}^{t} h(x)\,dx$, (averages $\overline{E_{\phi}(t)}, \;\overline{{\it\Gamma}_{\phi}(t)}$ are defined analogously). Within a rigorous treatment of the problem
it is straightforward to show that basis assumptions of the quantum theory guarantee that (see, e.g. \cite{ku-pra}),
\begin{equation}
\lim_{t \to \infty} {\it\Gamma}_{\phi}(t) = 0 \quad \text{and} \quad \lim_{t\to \infty}\overline{{\it\Gamma}_{\phi}(t)} =0. \label{limG=0}
\end{equation}
These results are rigorous. For $\overline{E_{\phi}(t)}$ one can show that $\lim_{t \to \infty}\overline{E_{\phi}(t)} = E_{\text{min}}$ (see \cite{Urbanowski:2008kra}).

The relations (\ref{h||2}), (\ref{h(t)}) are convenient when the density $\omega (E)$ is given and one wants to find the instantaneous energy $E_{\phi}(t)$ and decay rate ${\it\Gamma}_{\phi}(t)$: Inserting $\omega (E)$ into (\ref{a-spec}) one obtains the amplitude $A(t)$ and then using (\ref{h(t)}) one finds the $h(t)$ and thus $E_{\phi}(t)$ and ${\it\Gamma}_{\phi}(t)$. In the general case the density $\omega(E)$ posses properties analogous to the scattering amplitude, i.e., it can be decomposed into a threshold factor, a pole-function $P(E)$ with a simple pole and a smooth form factor $F(E)$. There is $\omega(E)= {\it\Theta}(E-E_\text{min})\,(E-E_\text{min})^{\alpha_{l}}\,P(E)\,F(E) $, where $\alpha_{l}$ depends on the angular momentum $l$ through $\alpha_{l} = \alpha + l$, \cite{Fonda:1978dk} (see equation (6.1) in \cite{Fonda:1978dk}), $0 \leq \alpha <1$) and ${\it\Theta}(E)$ is a step function: ${\it\Theta}(E) = 0\;\;\text{for}\;\; E \leq 0$ and ${\it\Theta}(E) = 1\;\;\text{for}\;\; E > 0$. The simplest choice is to take $\alpha = 0, l=0, F(E) = 1$ and to assume that $P(E)$ has a Breit--Wigner (BW) form of the energy distribution density. (The mentioned Breit--Wigner distribution was found when the cross-section of slow neutrons was analyzed \cite{Breit:1936zzb}). It turns out that the decay curves obtained in this simplest case are very similar in form to the curves calculated for the above described more general $\omega(E)$, (see \cite{Kelkar:2010qn} and analysis in \cite{Fonda:1978dk}). So to find the most typical properties of the decay process it is sufficient to make the relevant calculations for $\omega (E)$ modeled by the the Breit--Wigner distribution of the energy density:
$
\omega (E) \equiv \omega_{\text{BW}}(E) \stackrel{\text{def}}{=} \frac{N}{2\pi}\, {\it\Theta} (E - E_{\text{min}}) \
\frac{{\it\Gamma}_{0}}{(E-E_{0})^{2} + (\frac{{\it\Gamma}_{0}}{2})^{2}},
$
where $N$ is a normalization constant. The parameters $E_{0}$ and ${\it\Gamma}_{0}$ correspond to the energy of the system in the unstable state and its decay rate at the exponential (or canonical) regime of the decay process. $E_{\text{min}}$ is the minimal (the lowest) energy of the system. Inserting $\omega_\text{BW}(E)$ into formula (\ref{a-spec}) for the amplitude $A(t)$ after some algebra one finds that
\begin{equation}
A(t) = A(t - t_{0}) = \frac{N}{2\pi}\,
e^{\textstyle{ - \frac{i}{\hbar} E_{0}t }}\, I_{\beta}\left(\frac{{\it\Gamma}_{0} (t-t_{0})}{\hbar}\right), \label{I(t)a}
\end{equation}
where
\begin{equation}
I_{\beta}(\tau) \stackrel{\rm def}{=}\int_{-\beta}^{\infty}
 \frac{1}{\eta^{2}
+ \frac{1}{4}}\, e^{\textstyle{ -i\eta\tau}}\,d\eta. \label{I(t)}
\end{equation}
Here $\tau = \frac{{\it\Gamma}_{0} (t-t_{0})}{\hbar} \equiv \frac{t-t_{0}}{\tau_{0}}$, $\tau_{0}$ is the lifetime, $\tau_{0} = \frac{\hbar}{{\it\Gamma}_{0}}$, and $\beta = \frac{E_{0} - E_{min}}{{\it\Gamma}_{0}}\,>\,0$. (The integral $I_{\beta}(\tau)$ can be expressed in terms of the integral--exponential function \cite{sluis,Urbanowski:2006mw,ku-2009} (for a definition, see \cite{olver,abramowitz}).

Note that the more convenient is to use $t'=(t - t_{0})$ in (\ref{I(t)a}), (\ref{I(t)}) or (\ref{a-spec}) and in formula of this type, or to assume that $t_{0} = 0$ in all formulae of this type, because this does not changes the results of calculations but makes them easier. So from this point we will assume that $t_{0} = 0$.

Next using this $A(t)$ given by relations (\ref{I(t)a}), (\ref{I(t)}) and the relation (\ref{h(t)}) defining the effective Hamiltonian $h_{\phi}(t)$ one finds that within the Breit--Wigner (BW) model considered
\begin{equation}
h(t) = E_{0} + {\it\Gamma}_{0}\,\frac{J_{\beta}(\frac{{\it\Gamma}_{0} t}{\hbar})}{I_{\beta}(\frac{{\it\Gamma}_{0} t}{\hbar})}, \label{h(t)-1}
\end{equation}
where
\begin{equation}
J_{\beta}(\tau) = \int_{- \beta}^{\infty}\,\frac{x}{x^{2} + \frac{1}{4}}\,e^{\textstyle{-ix\tau}}\,dx. \label{J-R}
\end{equation}
Working within the BW model and using $J_{\beta}(\tau)$ one should remember that $J_{\beta}(0)$ is undefined ($\lim_{\tau \to 0} \,J_{\beta}(\tau) = \infty$). Simply within the model defined by the Breit-Wigner distribution of the energy density, $\omega_\text{BW}(E)$, the expectation value of $\mathfrak{H}$, that is $\langle \phi|\mathfrak{H}|\phi \rangle $ is not finite. So all the consideration based on the use of $J_{\beta}(\tau)$ are valid only for $\tau > 0$.

It is relatively simply to find analytical form of $J_{\beta}(\tau)$ using the following identity
\begin{equation}
J_{\beta}(\tau) \equiv i\frac{\partial I_{\beta} (\tau)}{\partial \tau} , \label{J-R-eq}
\end{equation}

We need to know the energy of the system in the unstable state $|\phi\rangle$ considered and its decay rate. The instantaneous energy $E_{\phi}(t)$ of the system in the unstable state $|\phi\rangle$ has the following form within the BW model considered
\begin{equation}
E_{\phi}(t) = \Re\,[h(t)] = E_{0} + {\it\Gamma}_{0}\,\Re\left[\frac{J_{\beta}(\frac{{\it\Gamma}_{0} t}{\hbar})}{I_{\beta}(\frac{{\it\Gamma}_{0} t}{\hbar})}\right], \label{h(t)-2}
\end{equation}
whereas the instantaneous decay rate looks as follows
\begin{multline}
{\it\Gamma}_{\phi}(\tau) = -2\Im\,[h(t)] = -\,2\,{\it\Gamma}_{0}\,\Im\left[\frac{J_{\beta}(\tau)}{I_{\beta}(\tau)}\right]\\ \equiv
 -\,2\,{\it\Gamma}_{0}\,\Im\left[\frac{J_{\beta}(\frac{{\it\Gamma}_{0} t}{\hbar})}{I_{\beta}(\frac{{\it\Gamma}_{0} t}{\hbar})}\right]. \label{G}
\end{multline}

 It is relatively simple to find asymptotic expressions $I_{\beta}{\tau}$ and $J_{\beta}(\tau)$ for $\tau \to \infty$ directly from (\ref{I(t)}) and (\ref{J-R}) using , e.g., the method of the integration by parts. We have for $\tau \to \infty$
\begin{multline}
I_{\beta}(\tau) \simeq \frac{i}{\tau}\,\frac{e^{\textstyle{i\beta \tau}}}{\beta^{2} + \frac{1}{4}}\,\\
\Big\{-1 + \frac{2 \beta}{\beta^{2} + \frac{1}{4}}\,\frac{i}{\tau} + \Big[\frac{2}{\beta^{2} + \frac{1}{4}} - \frac{8 \beta^{2}}{(\beta^{2} + \frac{1}{4})^{2}}\Big]\,\Big(\frac{i}{\tau}\Big)^{2}\,+\,\ldots \Big\} \label{I-as}
\end{multline}
and
\begin{multline}
J_{\beta}(\tau) \simeq \frac{i}{\tau}\,\frac{e^{\textstyle{i\beta \tau}}}{\beta^{2} + \frac{1}{4}}\,\\
\Big\{\beta +
\Big[1 - \frac{2 \beta^{2}}{\beta^{2} + \frac{1}{4}}\Big]\,\frac{i}{\tau} + \frac{\beta}{\beta^{2} + \frac{1}{4}}\Big[\frac{8 \beta^{2}}{\beta^{2} + \frac{1}{4}} - 6\Big]\,\Big(\frac{i}{\tau}\Big)^{2} + \ldots \Big\}. \label{J-as}
\end{multline}
These two last asymptotic expressions allows one to find for $\tau \to \infty$ the asymptotic form of the ratio $\frac{J_{\beta}(\tau)}{I_{\beta}(\tau)}$ used in relations
(\ref{h(t)-1}), (\ref{h(t)-2}) and (\ref{G}), which has much simpler form than asymptotic expansions for $I_{\beta}(\tau)$ and $J_{\beta}(\tau)$. One finds that for $\tau \to \infty$,
\begin{equation}
\frac{J_{\beta}(\tau)}{I_{\beta}(\tau)} \,\simeq \, - \,\beta \;-\;\frac{i}{\tau}\;-\; \frac{2 \beta}{\beta^{2} + \frac{1}{4}}\,\frac{1}{\tau^{2}} \,\,+ \ldots .\label{J-I-as}
\end{equation}
Starting from this asymptotic expression and formula (\ref{h(t)-2}) one finds, eg. that for $t \to \infty$,
\begin{equation}
{E_{\phi}(t)\vline}_{\,t \rightarrow \infty} \simeq { E}_{\text{min}}\, -\,2\,
\frac{ { E}_{0}\,-\,{E}_{min}}{ |\,h_{\phi}^{0}\,-\,{E}_{\text{min}} \,|^{\,2} } \;
\left(\frac{\hbar}{t} \right)^{2} ,\label{Re-h-as}
\end{equation}
where $h_{\phi}^{0} = E_{0} - \frac{i}{2}{\it\Gamma}_{0}$ and,
\begin{equation}
{{\it\Gamma}_{\phi}(t)\vline}_{\,t \rightarrow \infty} \simeq 2{\it\Gamma}_{0}\,\frac{1}{\tau} +\ldots = 2\,\frac{\hbar}{t} +\ldots \, ,\label{Im-h-as}
\end{equation}
These two last relations are valid for $t > T$, where $T$ denotes the cross-over time, i.e. the time when exponential and late time inverse power law contributions to the survival amplitude become comparable.

\section{Metastable dark energy with a decay law \\from Quantum Mechanics}

Note that the model described by eqs.~(\ref{s-2-1})--(\ref{s-2-2}) is the classical physics model and therefore it cannot applied directly when one would like to follow Krauss and Dent and to consider the decay of the dark energy density $\rho_\text{DE}(t)$ as the quantum decay process. For example, the late time effects discussed in \cite{Krauss:2007rx} can never occur in the SHSS model. The simplest way to extend models considered in \cite{Shafieloo:2016bpk} so that they might be used to describe the decay of $\rho_\text{DE}(t)$ as a quantum process seems to be a replacement of the classical decay rate $\Gamma$ in eqs.~(\ref{s-2-1}), (\ref{s-2-2}) by the decay rate ${\it\Gamma}_{\phi}(t)/\hbar$ appearing in quantum theoretical considerations. It is because the classical decay rate $\Gamma_\text{class} = \Gamma$ corresponds to the quantum physics decay rate ${\it\Gamma}_\text{quant} = {\it\Gamma}_{\phi}(t)$ divided by $\hbar$ (that is to ${\it\Gamma}_{\phi}(t)/\hbar$) and using ${\it\Gamma}_{\phi}(t)$ one can insert it into eq.~(\ref{s-2-2}) to obtain
\begin{equation}
\dot{\rho}_\text{DE}(t) \,=\,-\,\frac{1}{\hbar}\,{\it\Gamma}_{\phi}(t)\,\rho_\text{DE}(t), \label{s-2-2-h-q}
\end{equation}
instead of the classical fundamental equation of the radioactive decays theory. In fact this equation is a simple improvement of models discussed in \cite{Shafieloo:2016bpk}, and it can be considered as the use of quantum corrections in the models mentioned. In such a case eq. (\ref{s-2-1}) takes the following form
\begin{align}
\rho_\text{DE}(t) &= \rho_\text{DE}(t_{0})\,\times\,\exp\left[- \frac{t}{\hbar}\,\overline{{\it\Gamma}_{\phi}(t)} \right]\label{s-2-1-h2+s}\\
&\equiv \rho_\text{DE}(t_{0})\,\times\,\exp\left[- \frac{1}{\hbar}\int_{t_{0}}^{t}\,{\it\Gamma}_{\phi}(x) \, dx \right],\label{s-2-1-h2}
\end{align}
where ${\it\Gamma}_{\phi}(t)$ is given by formula (\ref{G}) and $\overline{{\it\Gamma}_{\phi}(t)} \stackrel{\rm def}{=} \frac{1}{t}\int_{t_{0}}^{t}\,{\it\Gamma}_{\phi}(x) dx $ is the average decay rate for the time interval $[0,t]$. These relations replacing eq.~(\ref{s-2-1}) contain quantum corrections connected with the use of the quantum theory decay rate.

Note that using the identity (\ref{G-equiv}) and the relation (\ref{I(t)a}) one rewrite the relation (\ref{s-2-1-h2}) as follows
\begin{equation}
\rho_\text{DE}(t) \,\equiv \, \frac{N^{2}}{4\pi^{2}} \;\rho_\text{DE}(t_{0})\, \,\left| I_{\beta} \left( \frac{{\it\Gamma}_{0}(t-t_{0}) }{\hbar} \right) \right|^{2}, \label{s-2-1-h2+A}
\end{equation}
which can make simpler numerical calculations.

Now in order to obtain analytical or numerical results having eqs.~(\ref{s-2-2-h-q})--(\ref{s-2-1-h2}) one needs a quantum mechanical model of the decay process, that is one needs $\omega (E)$ (see (\ref{a-spec}). We begin our considerations using the Breit-Wigner model analyzed in the previous section. Inserting ${\it\Gamma}_{\phi}(t)$ given by (\ref{G}) into eq.~(\ref{s-2-2-h-q}), or eqs.~(\ref{s-2-1-h2+s}), (\ref{s-2-1-h2}) we can analyze the decay process of $\rho_\text{DE}(t)$. One can notice that performing calculations, e.g. using the Breit--Wigner model, it is more convenient to use eq. (\ref{s-2-1-h2+A}) with $I_{\beta}(t)$ given by the relation (\ref{I(t)}) than using eqs. (\ref{s-2-1-h2+s}), (\ref{s-2-1-h2}) with ${\it\Gamma}_{\phi}(t)$ given by formula (\ref{G}).

Note that one of parameters appearing in quantum mechanical formula (\ref{G}) for ${\it\Gamma}_{\phi}(t)$ is ${\it\Gamma}_{0}$. This parameter can be eliminated if we notice that
$
\beta = \frac{E_{0} - E_{\text{min}}}{{\it\Gamma}_{0}} > 0.
$
Hence ${\it\Gamma}_{0} \equiv \frac{E_{0} - E_\text{min}}{\beta}$, and therefore one can rewrite (\ref{G}) as
\begin{equation}
{\it\Gamma}_{\phi}(\tau) = - 2\,\frac{E_{0} - E_\text{min}}{\beta}\,\Im\left[\frac{J_{\beta}(\tau)}{I_{\beta}(\tau)}\right], \label{G2}
\end{equation}
or,
\begin{equation}
{\it\Gamma}_{\phi}(\tau) = - 2\,\frac{\frac{E_{0}}{V_{0}} - \frac{E_{\text{min}}}{V_{0}}}{\beta}\,V_{0}\,\;\Im\left[\frac{J_{\beta}(\tau)}{I_{\beta}(\tau)}\right], \label{G2a}
\end{equation}
where $V_{0}$ is the volume of the system at $t=t_{0}$.
We have $\frac{E_{0}}{V_{0}} = \rho_\text{DE}^{qft} \stackrel{\rm def}{=} \rho_{DE}^{0}$ and
$\frac{E_{\text{min}}}{V_{0}} = \rho_\text{bare}$, (where $\rho_{DE}^{qft}$ is the energy density calculated using quantum field theory methods), so eq. (\ref{G2a}) can be rewritten as follows
\begin{equation}
{\it\Gamma}_{\phi}(\tau) = - 2\,\frac{\rho_\text{DE}^{0}- \rho_\text{bare}}{\beta}\,V_{0}\,\;\Im\left[\frac{J_{\beta}(\tau)}{I_{\beta}(\tau)}\right], \label{G2aa}
\end{equation}
The parameter $\tau$ used in (\ref{G2})--(\ref{G2aa}) denotes time $t$ measured in lifetimes as it was mentioned after formula (\ref{I(t)}): $\tau = \frac{t}{\tau_{0}}$. Using the parameter $\beta$ the lifetime $\tau_{0}$ can be expressed as follows: $\tau_{0} = \frac{\beta}{\rho_\text{DE}^{0} - \rho_\text{bare}}\,\frac{\hbar}{V_{0}}$.

The asymptotic form (\ref{Im-h-as}) indicates one of main differences between SHSS model and our improvement of this model. Namely, from eq. (\ref{s-2-1}) it follows that
\begin{equation}
\lim_{t \to \infty} \rho_\text{DE}(t) = 0. \label{rho=0}
\end{equation}
From (\ref{s-2-1}) one sees that $\rho_\text{DE}(t)$ is exponentially decreasing function of time.

It is interesting to consider a more general form of the energy density,
\begin{equation}
\tilde{\rho}_\text{DE}(t) = \rho_\text{DE}(t) - \rho_\text{bare},\label{rho-tilde}
\end{equation}
where $\rho_{bare} = \text{const}$ is the minimal value of the dark energy density. Inserting the density $\tilde{\rho}_{DE}(t)$ into eq. (\ref{s-2-1}) one concludes that $\rho_\text{DE}(t)$ tends to $\rho_\text{bare}$ exponentially fast as $t\to\infty$.

Let us see now what happens when to insert $\tilde{\rho}_\text{DE}(t)$ into our eq. (\ref{s-2-2-h-q}) and to consider only the asymptotic behavior of $\rho_\text{DE}(t)$ for times $t \geq T_{0} \gg T$. In such a case inserting the late time asymptotic expression (\ref{Im-h-as}) into eq. (\ref{s-2-2-h-q}) one finds for very late times $ t > T_{0}$ that
\begin{equation}
\ln\,\frac{\tilde{\rho}_\text{DE}(t)}{\tilde{\rho}_\text{DE}(T_{0})} = \ln\,\Big(\frac{t}{T_{0}}\Big)^{-2}, \label{rho-quant-as}
\end{equation}
that is that for $t > T_{0} \gg T$,
\begin{equation}
\rho_\text{DE}(t) \simeq \rho_\text{bare} + D\,\frac{1}{t^{2}}, \label{rho-as}
\end{equation}
where $D = const$.
Note that the same result follows directly from (\ref{s-2-1-h2+A}) when one inserts there $A(t)$ given by formula (\ref{I(t)a}) and uses asymptotic expression (\ref{I-as}) for $I_{\beta}(\tau)$, which shows that our approach is self-consistent. The result (\ref{rho-as}) means that quantum corrections does not allow $\rho_\text{DE}(t)$ tend to $\rho_\text{bare}$ exponentially fast when $t \to \infty$ but there must be that $\rho_\text{DE}(t)$ tends to $\rho_\text{bare}$ as $1/t^{2}$, for $t \to \infty$, which is in the full agreement with our earlier results presented, e.g, in \cite{Urbanowski:2011zz,Urbanowski:2012pka,Szydlowski:2015kqa,Szydlowski:2015bwa,Urbanowski:2016pks}. So in fact, as one can see, the SSHS model is the classical physics approximation of the model discussed in our papers mentioned, where cosmological parametrization resulting from the quantum mechanical treatment of unstable systems was used.

\section{Cosmological equations}

We introduce our model as the covariant theory with the interaction term \cite{Josset:2016vrq}. We consider the flat cosmological model (the constant curvature is equal zero).

The total density of energy consists of the baryonic matter $\rho_\text{B}$, the dark matter $\rho_\text{DM}$ and the dark energy $\rho_\text{DE}$. We assume, for the baryonic matter and the dark matter, the equation of state for dust ($p_\text{B}(\rho_\text{B})=0$ and $p_\text{DM}(\rho_\text{DM})=0$). And we consider the equation of state for the dark energy as $p_\text{DE}(\rho_\text{DE})=-\rho_\text{DE}$.

The cosmological equations such as the Friedmann and acceleration equations are found by the variation action by the metric $g_{\mu \nu}$ \cite{Josset:2016vrq}. In consequence we get the following equations
\begin{equation}
3 H^2=3 \frac{\dot a}{a}^2=\rho_\text{tot}=\rho_\text{B}+\rho_\text{DM}+\rho_\text{DE} \label{fried}
\end{equation}
and
\begin{equation}
\frac{\ddot a}{a}=-\frac{1}{6}(\rho_\text{tot}+3 p_\text{tot}(\rho_\text{tot}))=\rho_\text{B}+\rho_\text{DM}-2\rho_\text{DE},\label{acceleration}
\end{equation}
where $H=\frac{\dot a}{a}$ is the Hubble function. Here, we assume $8\pi G=c=1$.

Equations (\ref{fried}) and (\ref{acceleration}) give the conservation equation in the following form
\begin{equation}
\dot\rho_\text{tot}=-3H (\rho_\text{tot}+p_\text{tot}(\rho_\text{tot}))
\end{equation}
or in the equivalent form
\begin{equation}
\dot\rho_\text{M}=-3H \rho_\text{M}-\dot{\rho}_\text{DE},\label{conservation}
\end{equation}
where $\rho_\text{M}=\rho_\text{B}+\rho_\text{DM}$.

Let $Q$ denotes the interaction term. Equation~(\ref{conservation}) can be rewritten as
\begin{multline}
\dot\rho_\text{b} = -3H \rho_\text{B}, \text{ }\dot\rho_\text{DM} = -3H \rho_\text{DM}+Q \, \text{and} \,
\dot\rho_\text{DE} =-Q. \label{darkenergy}
\end{multline}
If $Q>0$ then the energy flows from the dark energy sector to the dark matter sector. If $Q<0$ then the energy flows from the dark matter sector to the dark energy sector.

\begin{figure}
	\centering
	\includegraphics[width=1.\linewidth]{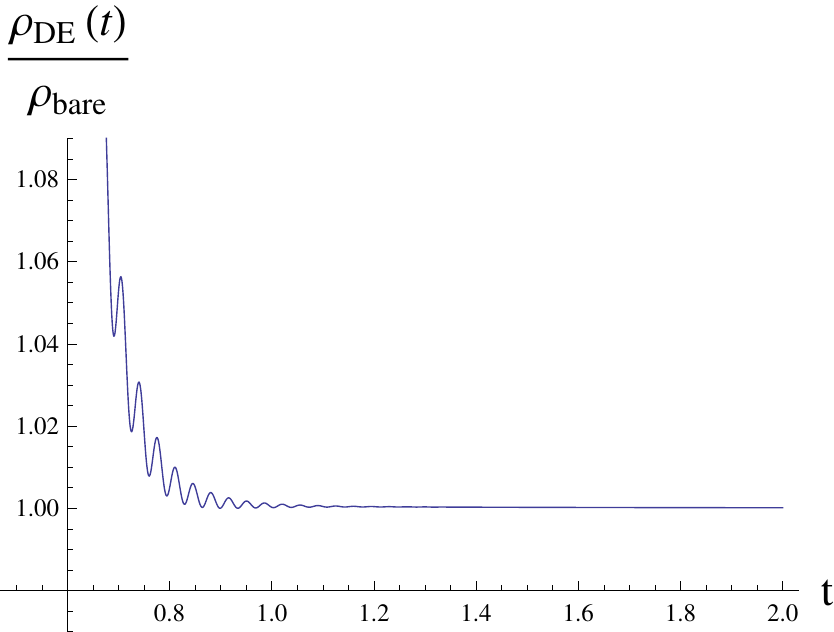}
	\caption{The dependence $\rho_\text{DE}(t)$ (from formula (\ref{darkenergy2})). For illustration we put $\beta=800$, ${\it\Gamma}_0=20 \hbar$ and $\epsilon=1000 \rho_\text{bare}$. The qualitative behaviour of $\rho_\text{DE}$ does not depend on $\epsilon$. The units of time $t$ are determined by a choice of units of ${\it\Gamma}_0$ because $\frac{{\it\Gamma}_0 t}{\hbar}$ are dimensionless.
}
	\label{fig:fig1}
\end{figure}

\begin{figure}
	\centering
	\includegraphics[width=1.0\linewidth]{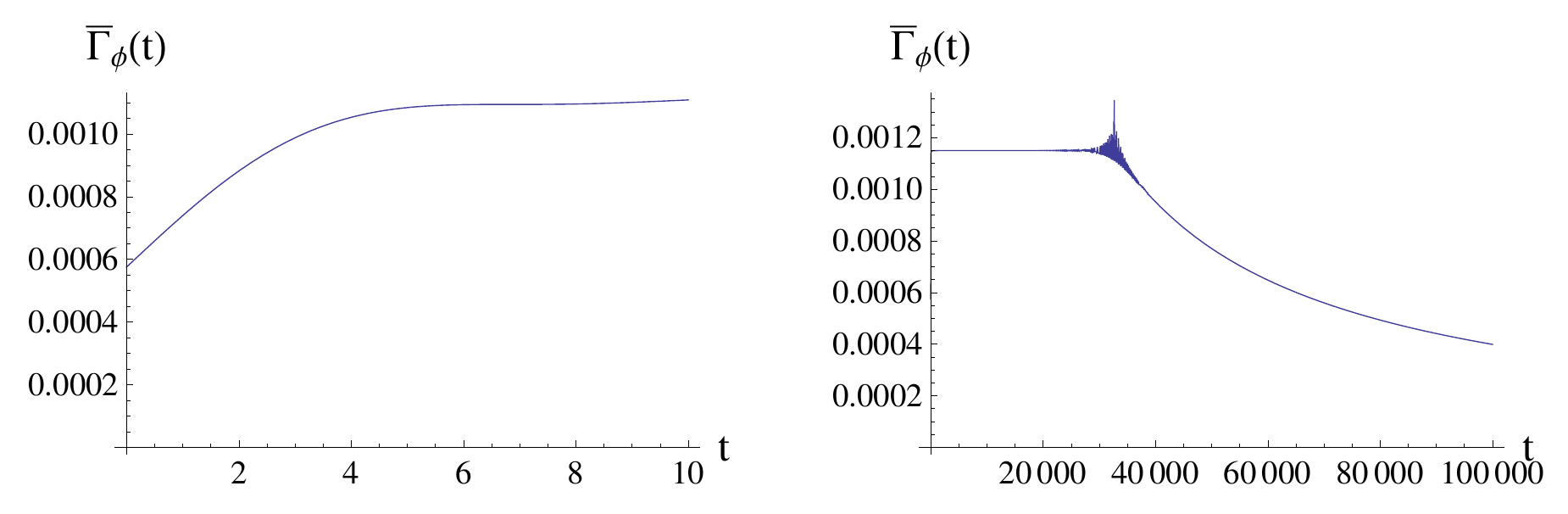}
\caption{The dependence $\bar{\it\Gamma}_\phi(t)$ for the best fit values (see table~\ref{table:1}). The left panel presents the evolution of $\bar\Gamma_\phi(t)$ for the early universe and the present epoch. The right panel presents evolution of $\bar{\it\Gamma}_\phi(t)$ for the late time universe. The cosmological time $t$ is expressed in $\frac{\text{s}\times \text{Mpc}}{\text{100 km}}$. In these units, the age of the Universe is equal 1.41 $\frac{\text{s}\times \text{Mpc}}{\text{100 km}}$.}
	\label{fig:fig2}
\end{figure}

Figure~\ref{fig:fig1} shows the diagrams of the evolution of $\rho_\text{DE}(t)$. Note that the oscillatory phase appears appears in the evolution of $\rho_\text{DE}(t)$.
Figure~\ref{fig:fig2} presents the evolution of the $\it{\bar\Gamma}_{\phi}(t)$. At the initial period we obtain a logistic-type decay of dark energy. Thy period when $\bar{\Gamma}_{\phi}(t)$ grows up to a plateau is characteristic for the so called the Zeno time \cite{Chiu:1977ds}. It increases slowly about $0.0004$ (the slope of this curve is $0.0001$) with the cosmic time $t$ in the interval $(0,4)$. Then, in the interval $(4, 30000)$ becomes strictly constant. This behavior justifies a radioactive approximation given in ref.~\cite{Shafieloo:2016bpk}. For the late time $\it{\bar{\it\Gamma}}_{\phi}(t)$ approaches to zero. \\

Using
(\ref{s-2-1-h2+A}) we get the final formula for $\rho_\text{DE}(t)$
\begin{equation}
\rho_\text{DE}(t)=\rho_{\text{bare}}+\epsilon \left|I_\beta\left(\frac{{\it\Gamma}_0 t}{\hbar}\right)\right|^2,\label{darkenergy2}
\end{equation}
where $\epsilon \equiv \epsilon (\beta) = \frac{\rho_\text{DE}(0)-\rho_{\text{bare}}}{\left|I_\beta\left(0 \right)\right|^2}$ measures the deviation from the $\Lambda$CDM model ($I_{\beta}(0) \equiv \frac{2\pi}{N} = \pi + 2 \arctan(2\beta)$ and $\beta >0$).

\begin{figure}
	\centering
	\includegraphics[width=1.\linewidth]{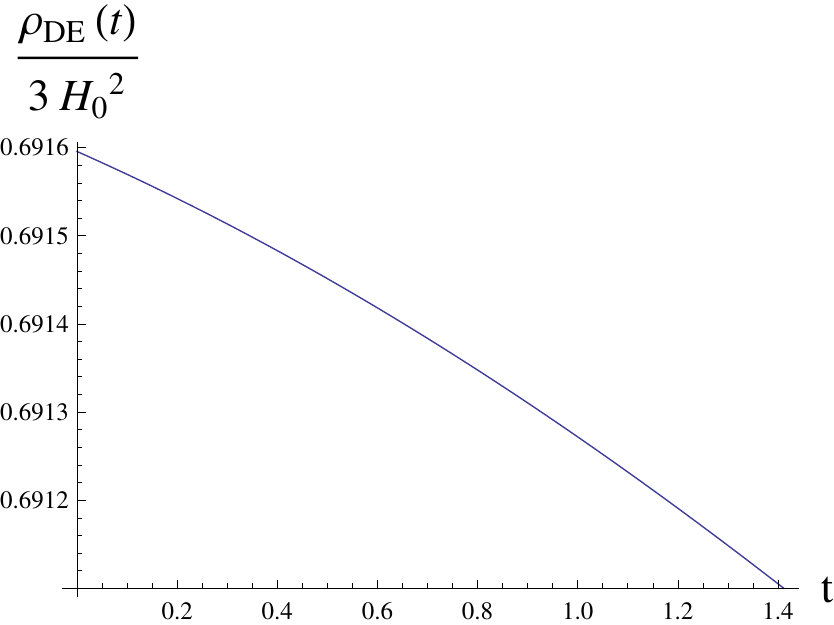}
	\caption{The dependence $\rho_\text{DE}(t)$ (from formula (\ref{darkenergy2})) for the best fit value of model parameter (see Table \ref{table:1}). The cosmological time $t$ is expressed in $\frac{\text{s}\times\text{Mpc}}{100\text{ km}}$. The present epoch is for $t=1.41 \frac{\text{s}\times\text{Mpc}}{100\text{ km}}$. Note that, in the Planck epoch, the value of $\frac{\rho_\text{DE}(t_\text{Pl})}{3H_0^2}$ is equal 0.6916.}
	\label{fig:fig5}
\end{figure}

The canonical scaling law for cold dark matter should be modified. In this case
\begin{equation}
\rho_\text{DM}=\rho_{\text{DM},0}a^{-3+\delta},
\end{equation}
where $\delta=\frac{1}{\ln{a}}\int \frac{Q}{H\rho_\text{DM}}d\ln a$. The evolution of $\delta$, for example values, is presented in figure~\ref{fig:fig4} and for the best fit values is presented in figure~\ref{fig:fig5}.

\begin{figure}
	\centering
	\includegraphics[width=1.\linewidth]{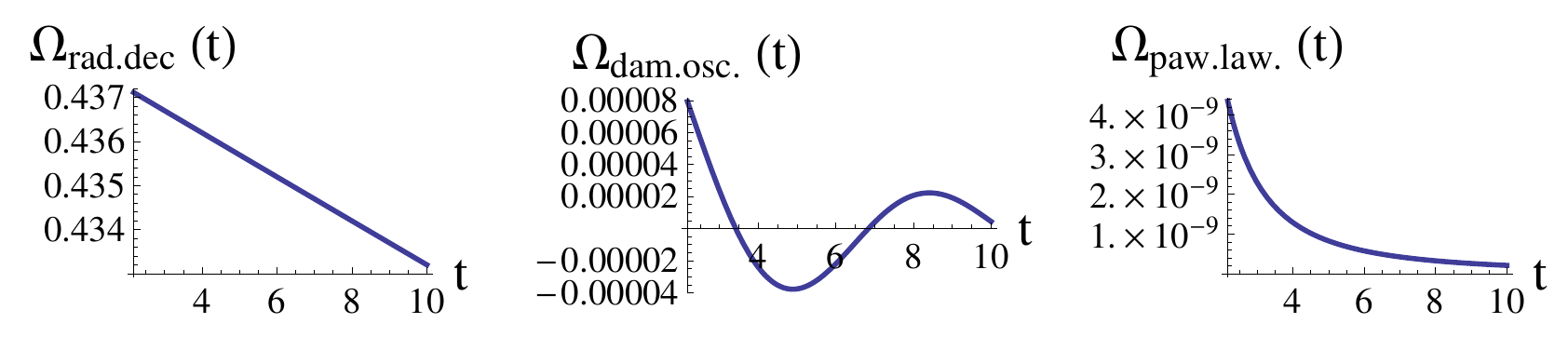}
	\caption{The dependence $\Omega_\text{rad.dec.}$, $\Omega_\text{dam.osc.}$, $\Omega_\text{pow.law}$ with respect to the cosmological time $t$ for the best fit value of model parameter (see Table \ref{table:1}). The cosmological time $t$ is expressed in $\frac{\text{s}\times\text{Mpc}}{100\text{ km}}$. In these units, the present epoch is for $t=1.41 \frac{\text{s}\times\text{Mpc}}{100\text{ km}}$. Let us note that while the density parameters do not change practically during the cosmic evolution for the cases shown in left and central panels, the density parameters are lowering by many orders of magnitude for the case presented in the right panel \cite{Szydlowski:2015bwa}.}
	\label{fig:fig6}
\end{figure}

\begin{figure}
	\centering
	\includegraphics[width=1.\linewidth]{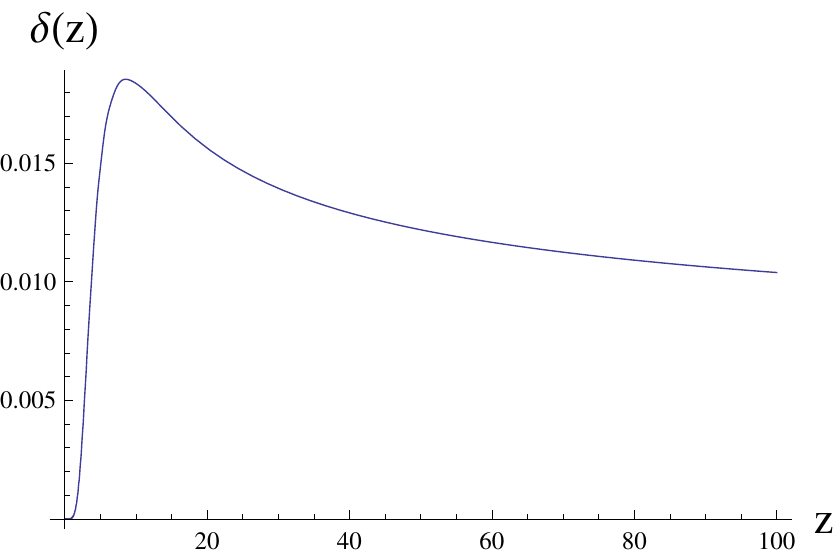}
	\caption{A diagram of the evolution of $\delta(z)$, where $z$ is redshift. For illustration we put $\beta=800$, ${\it\Gamma}_0=20 \hbar$ and $\epsilon=1000 \rho_\text{bare}$. Function $\delta(z)$ reaches the maximum for $z=z_{0}$, which is a solution of equation $\delta(z_{0})=\frac{Q(z_{0})}{H(z_{0})\rho_\text{DM}(z_{0})}$.}
	\label{fig:fig4}
\end{figure}

Assuming that $\beta >0$ one obtains for $t > t_{L} = \frac{\hbar}{\it{\Gamma}_{0}}\,\frac{2\beta}{\beta^{2} + \frac{1}{4}}$ (see \cite{sluis}) the approximation of (\ref{darkenergy2}) in the following form
\begin{multline}
\rho_\text{DE}(t)\approx\rho_\text{bare}+\\
\epsilon \left(4\pi^2 e^{-\frac{\it{\Gamma}_0}{\hbar}t}+\frac{4\pi e^{-\frac{\it{\Gamma}_0}{2\hbar}t}\sin\left(\beta\,\frac{\it{\Gamma}_0}{\hbar}t\right)}{\left(\frac{1}{4}+{\beta^2}\right)\frac{\it{\Gamma}_0}{\hbar}t}+
\frac{1}{\left(\left(\frac{1}{4}+{\beta^2}\right)\frac{\it{\Gamma}_0}{\hbar}t\right)^2}\right).\label{darkenergy3}
\end{multline}
For the best fit value (see Table \ref{table:1}) $t_L\approx 2 T_0$, where $T_0$ is the present age of the Universe.

From formula (\ref{darkenergy3}), it results that, for the late time, the behavior of dark energy can be described by the following formula
\begin{equation}
\rho_\text{DE}(t)\approx\rho_\text{bare}+\frac{\epsilon}{\left(\left(\frac{1}{4}+ {\beta^2}\right)\frac{\it{\Gamma}_0}{\hbar}\right)^2}\frac{1}{t^2}.\label{darkenergy4}
\end{equation}
This case was considered in \cite{Szydlowski:2015rga, Szydlowski:2015fya}.

If we use formula (\ref{darkenergy3}) in the Friedmann equation (\ref{fried}), we get
\begin{multline}
3 H^2=\rho_\text{tot}=\rho_\text{B}+\rho_\text{DM}+\rho_\text{bare}+\rho_\text{rad.dec.}\\ +\rho_\text{dam.osc.}+\rho_\text{pow.law},\label{friedmann}
\end{multline}
where $\rho_\text{rad.dec.}=4\pi^2 \epsilon e^{-\frac{{\it\Gamma}_0}{\hbar}t}$ is the radioactive-like decay part of dark energy, $\rho_\text{dam.osc.}=\frac{4\pi \epsilon e^{-\frac{\it{\Gamma}_0}{2\hbar}t}\sin\left(\beta\frac{\it{\Gamma}_0}{\hbar}t\right)}{\left(\frac{1}{4}+\beta^2\right)\frac{\it{\Gamma}_0}{\hbar}t}$ represents the damping oscillations part of dark energy and $\rho_\text{pow.law}=\frac{\epsilon}{\left(\left(\frac{1}{4}+\beta^2\right)\frac{\Gamma_0}{\hbar}t\right)^2}$ represents the power law part of dark energy. Using dimensionless parameters $\Omega_i=\frac{\rho_i}{3H_0^2}$, where $H_0$ is the present value of the Hubble constant, formula (\ref{friedmann}) can be rewritten as
\begin{equation}
\frac{H^2}{H_0^2}=\Omega_\text{B}+\Omega_\text{DM}+\Omega_\text{bare}+\Omega_\text{rad.dec.}+\Omega_\text{dam.osc.}+\Omega_\text{pow.law}. \label{friedmann2}
\end{equation}

If the radioactive-like decay dominates then one can define e-folding time $\lambda$ and half life time $T_{1/2}=\lambda \ln 2= \frac{\hbar\ln 2}{\it{\Gamma}_0}$.

The evolution of $\Omega_\text{rad.dec.}$, $\Omega_\text{dam.osc.}$, $\Omega_\text{pow.law}$ with respect to time, for the best fit value (see Table~\ref{table:1}), is presented in Fig.~\ref{fig:fig6}.

In the moment when the period of the radioactive-like decay $T_\text{end rad.dec.}$ finishes then the value of $\rho_\text{rad.dec.}$ is equal to the value of $\rho_\text{dam.osc.}$. It gets us a condition
\begin{equation}
4\pi^2 \epsilon e^{-\frac{{\it\Gamma}_0}{\hbar}t}=\frac{4\pi \epsilon e^{-\frac{\it{\Gamma}_0}{2\hbar}t}\sin\left(\beta\frac{\it{\Gamma}_0}{\hbar}t\right)}{\left(\frac{1}{4}+\beta^2\right)\frac{\it{\Gamma}_0}{\hbar}t}
\end{equation}
or after simplifying
\begin{equation}
\pi e^{-\frac{{\it\Gamma}_0}{2\hbar}t}=\frac{\sin\left(\beta\frac{\it{\Gamma}_0}{\hbar}t\right)}{\left(\frac{1}{4}+\beta^2\right)\frac{\it{\Gamma}_0}{\hbar}t}.\label{tendrad}
\end{equation}
Equation (\ref{tendrad}) has infinitely many solutions but $T_\text{end rad.dec.}$ is equal to the least positive real solution of (\ref{tendrad}) because the period of the radioactive-like decay is before the period of the damping oscillation decay.

Searching for the value of $T_\text{end rad.dec.}$ can be simplified by using of the upper envelope of oscillations of $\rho_\text{dam.osc.}$, which is given by
\begin{equation}
e_\text{upper}(t)=\frac{4\pi \epsilon e^{-\frac{\it{\Gamma}_0}{2\hbar}t}}{\left(\frac{1}{4}+\beta^2\right)\frac{\it{\Gamma}_0}{\hbar}t}.\label{envelope}
\end{equation}
Then we get an approximation of Eq. (\ref{tendrad}) in the form $\rho_\text{rad.dec.}=e_\text{upper}$ or after simplifying
\begin{equation}
\pi e^{-\frac{{\it\Gamma}_0}{2\hbar}t}=\frac{1}{\left(\frac{1}{4}+\beta^2\right)\frac{\it{\Gamma}_0}{\hbar}t}.\label{tendrad2}
\end{equation}
The solution of Eq. (\ref{tendrad2}) gives us the approximated value of $T_\text{end rad.dec.}$.

Note that solution of eq. (\ref{tendrad2}) cannot be less than the value of $T_\text{end rad.dec.}$ subtracted the value of one period of oscillation of $\rho_\text{dam.osc.}$ (i.e., $T_\text{dam.osc.}=\frac{2\pi\hbar}{\beta \it{\Gamma}_0}$) and cannot be greater than the value of $T_\text{end rad.dec.}$. In consequence for $\beta>29$, the error of the approximation is less than 1\%.

\begin{figure}
	\centering
	\includegraphics[width=1.\linewidth]{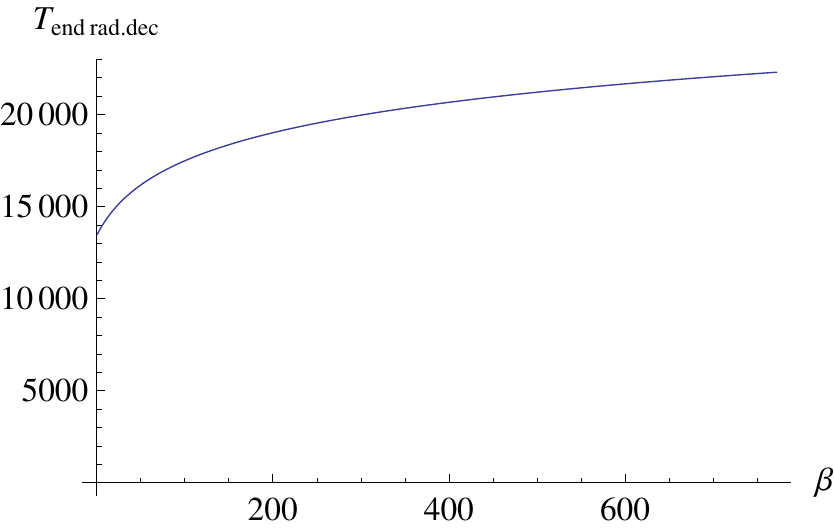}
	\caption{A diagram presents a dependence $T_\text{end rad.dec.}(\beta)$ for $\beta>29$. For illustration we put the best fit value of ${\it\Gamma}_0$ (see Table \ref{table:1}). Values of $T_\text{end rad.dec.}$ are expressed in the present age of the Universe. We get this diagram from numerical analysis of Eq. (\ref{tendrad2}).}
	\label{fig:fig8}
\end{figure}

For the best fit values (see Table \ref{table:1}), Eq. (\ref{tendrad2}) gives $T_\text{end rad.dec.}=22296\times T_0$, where $T_0$ is the present age of the Universe. The dependence $T_\text{end rad.dec.}(\beta)$ is presented in Fig.~\ref{fig:fig8}.

\section{Statistical analysis}

In our statistical analysis, it was used the following astronomical data: supernovae of type Ia (SNIa) (Union 2.1 dataset \cite{Suzuki:2011hu}), BAO data (Sloan Digital Sky Survey Release 7 (SDSS DR7) dataset at $z = 0.275$
\cite{Percival:2009xn}, 6dF Galaxy Redshift Survey measurements at redshift $z = 0.1$ \cite{Beutler:2011hx}, and WiggleZ measurements at redshift $z = 0.44, 0.60, 0.73$
\cite{Blake:2012pj}), measurements of the Hubble parameter $H(z)$ of galaxies
\cite{Simon:2004tf,Stern:2009ep,Moresco:2012jh}, the Alcock-Paczynski test (AP)\cite{Alcock:1979mp,Lopez-Corredoira:2013lca} (data from
\cite{Sutter:2012tf,Blake:2011ep,Ross:2006me,Marinoni:2010yoa,daAngela:2005gk,Outram:2003ew,Anderson:2012sa,Paris:2012iw,Schneider:2010hm}.) and measurements of CMB by Planck \cite{Ade:2015rim}. The formula for the likelihood function is given by 
\begin{equation}L_{\text{tot}} = L_{\text{SNIa}} L_{\text{BAO}} L_{\text{AP}}
	L_{H(z)} L_{\text{CMB}}.
\end{equation}
	
The likelihood function for SNIa has the form 
\begin{equation}
L_{\text{SNIa}} = \exp\left[-\frac{1}{2} [A - B^2/C
 + \log(C/(2 \pi))]\right],
\end{equation}
where $A=
(\mathbf{\mu}^{\text{obs}}-\mathbf{\mu}^{\text{th}})\mathbb{C}^{-1}(\mathbf{\mu}^{\text{obs}}-\mathbf{\mu}^{\text{th}})$,
$B = \mathbb{C}^{-1}(\mathbf{\mu}^{\text{obs}}-\mathbf{\mu}^{\text{th}})$,
$C=\text{Tr}\, \mathbb{C}^{-1}$ and $\mathbb{C}$ is a covariance matrix for SNIa, $\mu^{\text{obs}}$ is the observer distance modulus and $\mu^{\text{th}}$ is the theoretical distance modulus.

The likelihood function for BAO is described by the formula 
\begin{equation}
L_{\text{BAO}} = \exp\left[-
\frac{1}{2}\left(\mathbf{d}^{\text{obs}}-\frac{r_s(z_d)}{D_V(\mathbf{z})}\right)\mathbb{C}^{-1}\left(\mathbf{d}^{\text{obs}}-\frac{r_s(z_d)}{D_V(\mathbf{z})}\right)\right],
\end{equation}
where $r_s(z_d)$ is the sound horizon at the drag epoch \cite{Hu:1995en,Eisenstein:1997ik}. 

The likelihood function
\begin{equation}
L_{H(z)} = \exp\left[-\frac{1}{2} \sum_{i=1}^{N} \left
(\frac{H(z_i)^{\text{obs}}-H(z_i)^{\text{th}}}{\sigma_i
}\right)^2\right]
\end{equation}
is for measurements of the Hubble parameter $H(z)$ of galaxies. 

The likelihood function for AP is given by
\begin{equation}
L_{AP(z)} = \exp\left[-\frac{1}{2} \sum_{i=1}^{N} \left
(\frac{AP(z_i)^{\text{obs}}-AP(z_i)^{\text{th}}}{\sigma_i
}\right)^2]\right],
\end{equation}
where $AP(z)^{\text{th}} \equiv \frac{H(z)}{z} \int_{0}^{z}
\frac{dz'}{H(z')}$
and $AP(z_i)^{\text{obs}}$ are observational data. The likelihood function for CMB is given by 
\begin{equation} 
  L_{\text{CMB}} = \exp\left[- \frac{1}{2}
(\mathbf{x}^{\text{th}}-\mathbf{x}^{\text{obs}})
\mathbb{C}^{-1} (\mathbf{x}^{\text{th}}-\mathbf{x}^{\text{obs}})\right],
\end{equation}
where $\mathbb{C}$ is the covariance matrix with the errors, $\mathbf{x}$ is a vector
of the acoustic scale $l_{A}$, the shift parameter $R$ and $\Omega_{b}h^2$ where $
l_A = \frac{\pi}{r_s(z^{*})} c \int_{0}^{z^{*}} \frac{dz'}{H(z')}$ and $
R = \sqrt{\Omega_{\text{m,0}} H_0^2} \int_{0}^{z^{*}} \frac{dz'}{H(z')}$, where $z^{*}$ is the redshift of the epoch of the recombination \cite{Hu:1995en}.

\begin{figure}
   \includegraphics[scale=1.]{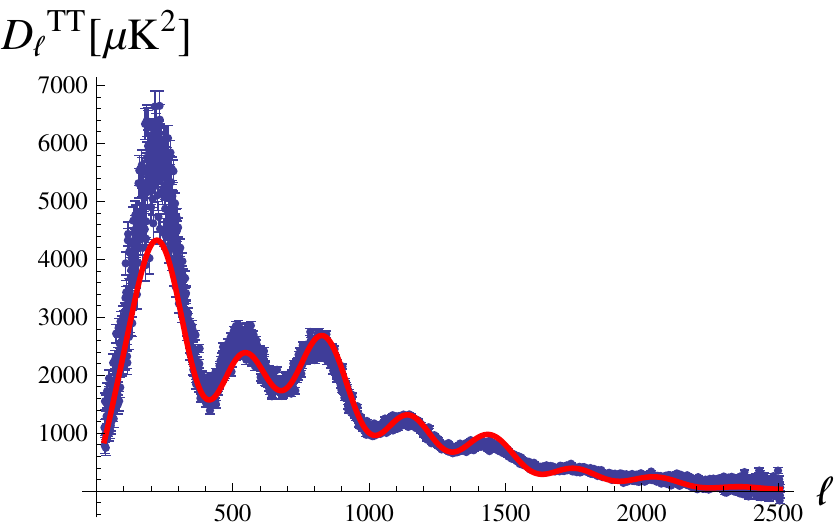}
   \caption{Diagram of the temperature power spectrum of CMB for the best fit values (red line). The error bars from the Planck data are presented by blue color.}
    \label{fig:fig3}
\end{figure}

\begin{table}
\caption{The best fit and errors for the estimated model with $\alpha$ from the interval (00.0, 0.033), $\Gamma_0/\hbar$ from the interval $(0.00 \frac{\text{100 km}}{\text{s}\times \text{Mpc}}, 0.036 \frac{\text{100 km}}{\text{s}\times \text{Mpc}})$ and $\epsilon/3H_0^2$ from the interval (0.00, 0.0175). We assumed that $\Omega_{\text{b},0}=0.048468$, $H_0=67.74 \frac{\text{km}}{\text{s}\times \text{Mpc}} $ and $\Omega_{\text{m},0}=0.3089$. In the table, the values of ${\it\Gamma}_0/\hbar$ are expressed in $\frac{\text{100 km}}{\text{s}\times \text{Mpc}}$. Because $\alpha=\frac{1}{1+\beta}$, the best fit value of $\beta$ parameter is equal 799.}
    \label{table:1}
    \begin{center}
        \begin{tabular}{llll} \hline
            parameter & best fit & $ 68\% $ CL & $ 95\% $ CL \\ \hline \hline
            $\alpha$ & 0.00125& $\begin{array}{c}
            +0.00104 \\ -0.00125
            \end{array}$ & $\begin{array}{c}
            +0.01777 \\ -0.00125
            \end{array}$ \\ \hline
            ${\it\Gamma}_0/\hbar$ & 0.00115 & $\begin{array}{c}
            +0.00209 \\ -0.00115
            \end{array}$ & $\begin{array}{c}
            +0.2123 \\ -0.00115
            \end{array}$ \\ \hline
            $\epsilon/3H_0^2$ & 0.0111 & $\begin{array}{c}
            +0.0064 \\ -0.0083
            \end{array}$ & $\begin{array}{c}
            +0.0064 \\ -0.0093
            \end{array}$ \\ \hline
        \end{tabular}
    \end{center}
\end{table}

In this paper, it was used our own code CosmoDarkBox in estimation of model parameter. Our code uses the Metropolis-Hastings algorithm \cite{Metropolis:1953am,Hastings:1970aa}.

In statistical analysis, we estimated three model parameters: $\alpha=\frac{1}{1+\beta}$, ${\it\Gamma}_0$, $\epsilon/3H_0^2$. Our statistical results are completed in table~\ref{table:1}. Diagram of the temperature power spectrum for the best fit values is presented in Fig.~\ref{fig:fig3}. Therefore the radioactive type of decay gives the most effective mechanism of decaying metastable dark energy. We estimated also that the decay half life time $T_{1/2}$ of dark energy is equal $8503 \text{ Gyr}\approx 616 \times T_0$, where $T_0$ is the present age of the Universe.

\section{Discussion and conclusions}

The main aim of this paper was to study the implication of a derived form of running dark energy. In our approach the formula for the parametrization of this dark energy is derived directly from the quantum mechanics rather than is postulated in a phenomenological way. The evolutional scenario of dark energy contain three different phases: a phase of radioactive-like decay in the early universe, a phase of damping oscillations and finally a phase of the power law type of decay.

We investigated a cosmological evolution caused by such variability of dark energy and matter. The dynamics of the model is governed by a cosmological dynamical system with an interacting term because the energy momentum tensor is not conserved in this case.

From investigation of variability of dark energy with the cosmological time, we demonstrated how the problem of the cosmological constant can be solved. We show that dark energy decays and then the canonical scaling law for cold dark matter $a^{-3}$ should be modified.

Using astronomical data we tested the model and obtain that it is in good agreement with the data. Our estimation also shows that the fraction of all components of dynamical dark energy in the whole dark energy is larger than contribution of the cosmological constant term.

In our model it is calculated that the $\Lambda$ term has a dynamical nature as a consequence of a decaying of the dark energy. In consequence the conservation of energy-momentum tensor (EMT) is violated. Recently Josset and Perez \cite{Josset:2016vrq} have demonstrated the model in which the violation of EMT can be achieved in the context of the unimodular gravity and how it leads to the emergence of the effective cosmological constant in Einstein’s equations. In our approach the violation of the conservation of EMT is rather a consequence of a quantum-mechanical nature of the metastable vacuum than a modification of the gravity theory.

In our approach the concrete form of decaying dark energy is derived directly from quantum mechanical consideration of unstable states. We obtain a more complex form of decaying dark energy in which we have found a radioactive type of its decaying. We also estimated the model parameters as well as fractions of three different forms of decaying: radioactive type, damping oscillating type and power-law type. From the astronomical data we obtain that the radioactive type of decay is favored and $44\%$ fraction of the energy budget of the Universe corresponds with a radioactive-like decay.

In our paper we investigate the second way of the decay of dark energy into dark matter from three different ways of dark energy decay considered by Shafieloo et al. \cite{Shafieloo:2016bpk}. They proposed a class of metastable dark energy models in which dark energy decays according to the radioactive law. They assumed a phenomenological form of the decay, studying observational constraints for the cosmological model. In our paper, it is derived directly from quantum mechanics as a result. Our results are complementary to their results because they justify the phenomenological choice of the exponential decay as a major mechanism of dark energy decay. Moreover, the our calculation of the decay half live is in agreement with Shafieloo et al.’s calculation. We obtain that the radioactive-like decay dominates up to $22296\times T_0$, where $T_0$ is the present age of the Universe. Our calculations show that the radioactive-like decay has only an intermediate character and will be replaced in the future evolution of the Universe by an oscillation decay and then decay of $1/t^2$ type.

The one of the differences between our approach and the theory developed by Shafieloo et al. is that they consider only decay of the dark energy into dark constituents assuming that the decay rate $\Gamma$ of the dark energy is constant and depends only on its internal composition. This last assumption is approximately true only if one considers decay processes as classical physics processes. The detailed analysis decay processes of quantum unstable systems shows that basic principles of the quantum theory does not allow them to be described by an exponential decay law at very late times as well as at initial stage of the decay process (see eg., \cite{Fonda:1978dk} and references therein, or \cite{ku-2017}) and that the decay law can be described by the exponentially decreasing function of time only at ``canonical decay regime'' of the decay process, that is at intermediate times (at times longer that initial stage of the evolution of the quantum unstable system and shorter than the crossover time $T$). These properties of quantum decay processes mean that in general the decay rate cannot be constant in time, $\Gamma ={\it\Gamma}(t) \neq \text{const}$ (see, eg., \cite{ku-pra,Giraldi:2015a,ku-2009,ku-2017}), and at the ``canonical decay'' stage ${\it\Gamma}(t) \simeq {\it\Gamma}_{0}$ to a very good approximation.

These properties of the decay rate was used in our paper. The advantage of the use of the decay rate following from the quantum properties of the decaying systems is that such an approach is able to describe correctly the initial stage of the dark energy decay process, and at very late times --- It is impossible to realize this within the approach used by by Shafieloo et al. What is more, the use $\Gamma = const$ may lead to the results which need not to be correct. The example of such situation is analysis performed in Appendix~A, Section A1, of the paper \cite{Shafieloo:2016bpk}, where the authors considered the case $\Gamma t \ll 1$ and then applied results obtained within such an assumption for the analysis of properties of their Model I. Namely, there are many reasons leading to the conclusion that the decay of the dark energy must be a quantum decay process (see discussion in Sec. 1) and that it can not be a classical physics process. So when one wants to describe the early stage of the decay process of the dark energy, which mathematically can be expressed by the assumption that $\Gamma t \ll 1$ one should not use the relation of the type (\ref{s-2-1}) but the relation
 \begin{equation}
 \rho_{\text{DE}}(t) = \rho_{\text{DE}}(0)\,|A(t)|^{2}, \label{rho-A}
 \end{equation}
resulting from the quantum mechanical treatment of the decay process. Instead of considering the relation of this type, authors of \cite{Shafieloo:2016bpk} used relation (\ref{s-2-1}) what leads to the formula (A1) in \cite{Shafieloo:2016bpk} for $\Gamma t \ll 1$, that is to
 \begin{equation}
 \rho_\text{DE} = \epsilon_{0}e^{\textstyle{- \Gamma t}}\,\simeq \,\epsilon_{0}(1 - \Gamma t), \label{rho-A1}
 \end{equation}
($\epsilon_{0}$ is defined in \cite{Shafieloo:2016bpk}), which is mathematically correct but it is not correct when one considers the decay of the dark energy as the quantum process. In the case of quantum decay process one should use the relation of the type (\ref{rho-A}) and the approximate form of $|A(t)|^{2}$ for very short times. There is in such a case (see, eg. \cite{Fonda:1978dk,ku-pra}),
 \begin{equation}
 |A(t)|^{2} \simeq 1 - d^{2}\,t^{2}, \;\;\, \text{for} \;\;\; t \to 0, \label{A-as-0}
 \end{equation}
 where $d = const$ and it does not depend on $\Gamma$. Therefore there should be 
 \begin{equation}
\rho_\text{DE} (t) \simeq \rho_\text{DE}(0)\,(1 - d^{2}\,t^{2}),\;\;\, \text{for}\;\;\; t \to 0, \label {rho-A-as-1}
\end{equation}
for shot times $t$, when the decay of the dark energy is a quantum decay process. The difference between relations (\ref{rho-A1}) (i.e., (A1) in \cite{Shafieloo:2016bpk}) and (\ref{rho-A-as-1}) is dramatic (and the the use $\epsilon_{0}$ in (\ref{rho-A1}) and $\rho_\text{DE}(0)$ in (\ref{rho-A-as-1}) is not the point). The problem is that authors of \cite{Shafieloo:2016bpk}) use their result (A1) (that is the above (\ref{rho-A1})) in formula (A2) and then all considerations related to their Model I in Section A I of the Appendix A are founded on relations (A1) and (A2). This means that these conclusions drawn in \cite{Shafieloo:2016bpk} (based on the analysis performed in Section A I of Appendix A) may not reflect real properties of decaying dark energy. In this place it should be noted that our analysis performed in this paper is free of this defect.

Note also that Shafieloo et al. \cite{Shafieloo:2016bpk} considered only the decays of the dark energy into dark components: dark matter and dark radiation, whereas we consider a general case (that is in our approach a decay of the dark energy into a visible baryonic matter is also admissible, which cannot be excluded in the light of the recently reported discovery of baryonic spindles linking galaxies \cite{Tanimura:2017ixt,deGraaff:2017byg,Fernini:2017}).

\end{document}